\documentclass[10pt,a4paper,twocolumn,english,pra,aps,showpacs,floatfix,groupedaddress,superscriptaddress]{revtex4}

\usepackage{graphicx}
\usepackage{epsfig}
\usepackage[english]{babel}
\usepackage{amsmath}
\usepackage{amssymb}
\usepackage{amsfonts}
\usepackage{longtable}
\usepackage{dcolumn}
\usepackage{bbm}

\begin{document}
\title{Numerical Analysis of Optimized Coherent Control Pulses}
\date{\rm\today}

\author{P. Karbach}
\affiliation{Lehrstuhl f\"{u}r Theoretische Physik I, 
Technische Universit\"{a}t Dortmund,
 Otto-Hahn Stra\ss{}e 4, 44221 Dortmund, Germany}

\author{S. Pasini}
\affiliation{Lehrstuhl f\"{u}r Theoretische Physik I, 
Technische Universit\"{a}t Dortmund,
 Otto-Hahn Stra\ss{}e 4, 44221 Dortmund, Germany}

\author{G. S. Uhrig}
\affiliation{Lehrstuhl f\"{u}r Theoretische Physik I, 
Technische Universit\"{a}t Dortmund,
 Otto-Hahn Stra\ss{}e 4, 44221 Dortmund, Germany}

\begin{abstract}
Numerically we simulate the effect of optimized coherent control pulses 
with a finite duration on a qubit in a bath of spins. The pulses of finite
duration are compared with ideal instantaneous pulses. In particular, we 
show that properly designed short pulses can approximate  ideal 
instantaneous pulses up to a certain order in the shortness of the  pulse.
We provide examples of such pulses, quantify the discrepancy from the 
ideal case and compare their effect for various ranges of the coupling 
constants.
\end{abstract}

\pacs{03.67.Lx, 03.67.Pp, 75.40.Mg, 76.60.-k}

\maketitle

\section{Introduction\label{intro}}
The coherent control of quantum systems continues to be a topic of great 
interest. The possibility of maintaining a spin in a coherent state is of 
extreme importance in fields of application like nuclear magnetic 
resonance (NMR) or the manipulation of quantum dots. In particular for 
quantum information processing, a long coherence time of the qubit is an 
indispensable prerequisite for its realization.

A quantum bit (henceforth: qubit) is a two-level system which is 
conveniently regarded as a spin $S=1/2$. Operations on qubits to change or
to correct their state are performed through 
quantum gates. Their effect on the density matrix of the qubit
can be described as a rotation in the Bloch sphere. 
Experimentally, they often can be obtained by the application of 
electromagnetic pulses. 
A 1-qubit gate is generally a single rotation about a given axis $\vec{a}$
in spin space. The angle of rotation classifies  the type of the pulse. 
For instance a $\pi$ pulses rotates the spin by 180$^\circ$. These pulses
 find a wide range of applications  in dynamical decoupling 
\cite{viola98,ban98,facch05,cappe06,witze07a,yao07,uhrig07} and in NMR
  \cite{haebe76,vande04}
where also $\pi/2$ pulses are crucial. In quantum information processing
the $\pi/2$ pulse in combination with a $\pi$ pulse
realizes the important Hadamard gate.

The idea of dynamical decoupling (DD)
\cite{viola98,ban98,facch05,cappe06,witze07a,yao07,uhrig07} has been 
developed from the spin echo technique in NMR 
\cite{hahn50,carr54,meibo58}. DD aims at decoupling the qubit from 
the environment  by means of the application of appropriate pulse 
sequences. From a theoretical point of view the topic has been widely 
studied and many different sequences of pulses have been proposed. Among 
these we recall the series of periodic equidistant $\pi$ pulses, called 
bang-bang control (BB) \cite{viola98,ban98}, the periodically iterated 
2-pulse sequence according to  Carr/Purcell and Meiboom/Gill (CPMG) 
\cite{carr54,meibo58,haebe76},
the concatenated sequence (CDD) proposed by Khodjasteh and Lidar 
\cite{khodj05,khodj07} as well as the fully optimized sequence
(UDD) derived by one of the authors \cite{uhrig07,uhrig08}.

Experimentally, the spin echo and the CPMG sequence are standard in 
NMR \cite{haebe76}. To our knowledge, other sequences have not yet
been tested. In realizations of qubits on the basis of semiconductor
technology so far only the spin echo technique has been implemented
 \cite{morto06,petta05,greil06a}. But computations for
quantum dot systems show that more elaborate pulse sequences
are very likely to be useful in suppressing decoherence, see for instance
\cite{yao07,witze07a,lee08a}.

Most theoretical examples (for exceptions see Ref. \cite{khodj05,khodj07}) reported so far have the limitation that the 
pulses are assumed to be ideal. This means that the pulses are considered 
to be instantaneous and infinitely strong in the sense of a 
$\delta$ peak. In this case, one is allowed to ignore the effect of the 
bath, inducing the decoherence, during the action of the pulse because 
the coupling to the bath is negligible relative to the amplitude of the 
$\delta$ pulse. Hence the rotation due to the pulse can be viewed to be 
completely separate from the free evolution of the system, qubit and bath,
 without pulse.

\begin{figure}
\begin{center}
     \includegraphics[width=\columnwidth]{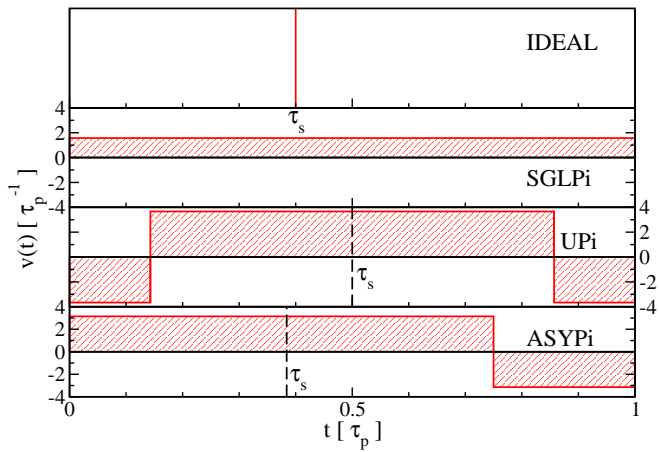}
\end{center}
\caption{(Color online) Examples of $\pi$ pulses implemented in the 
simulations. The ideal pulse is given by a $\delta$ peak operating at 
the instant $\tau_s$. SGLPi is the standard pulse of constant
amplitude without optimization of 
the pulse shape. For details see Table \ref{tab_pulses_pi}.}
 \label{pulsesPi}
\end{figure}

\begin{figure}
\begin{center}
     \includegraphics[width=\columnwidth]{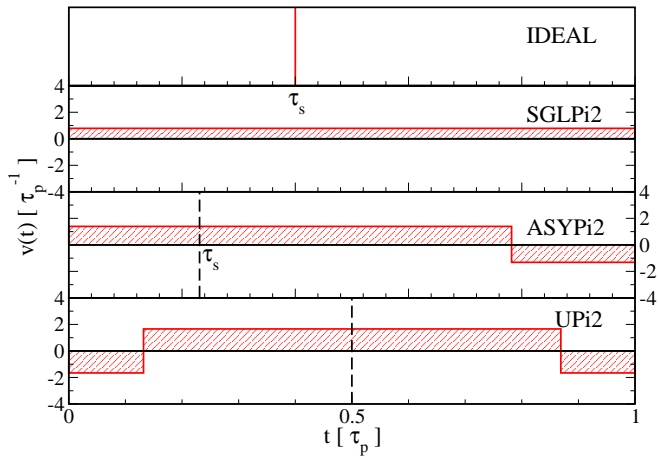}
\end{center}
\caption{(Color online) Examples of $\pi/2$ pulses implemented in the 
simulations. The ideal pulse is given by a $\delta$ peak operating at 
the instant $\tau_s$. SGLPi2 is the standard pulse of constant
amplitude without optimization of the pulse shape.  For details 
see Table \ref{tab_pulses_pi_halb}.}
\label{pulsesPi2}
\end{figure}

\begin{figure}
\begin{center}
     \includegraphics[width=\columnwidth]{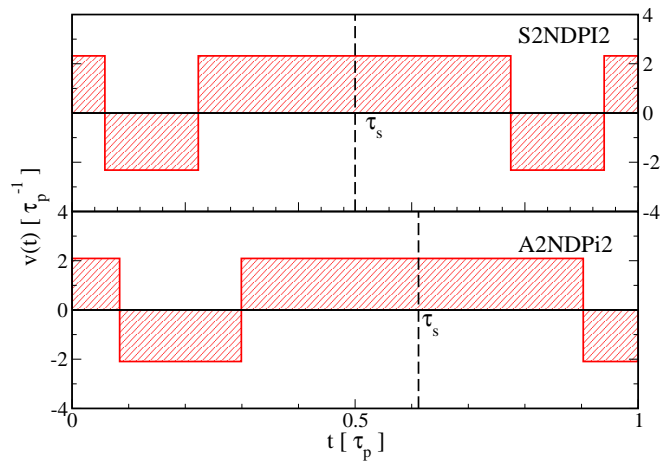}
\end{center}
\caption{(Color online)
Examples of $\pi/2$ pulses for which also some of 
the second order corrections vanish, namely $\eta_{21}=0$ and 
$\eta_{22}=0$ (see main text). For details see Table 
\ref{tab_pulses_pi_halb}.}
\label{pulsesPi2nd}
\end{figure}

If the pulse has a finite duration ($\tau_p$) so that its time of 
application is comparable with the characteristic time scales of the bath,
the separation between evolution due to the pulse and evolution of the 
undriven system is not valid anymore. If we suppose that the duration
$\tau_p$ is still small an expansion in $\tau_p$ about the limit
of a $\delta$ pulse is appropriate. The proposed  scenario \cite{pasin08a}
establishes an equivalence, up to corrections expanded in
a series in $\tau_p$, between the real pulse and an ideal
 $\delta$ pulse at some intermediate instant $\tau_s$ with 
$0<\tau_s<\tau_p$, see Figs.\ \ref{pulsesPi}, \ref{pulsesPi2}, and  
\ref{pulsesPi2nd}. Before and after the ersatz pulse at $\tau_s$, the 
free evolution of the system, qubit and bath, without pulse takes place.

The corrections expanded in powers of $\tau_p$
depend also on the shape of the pulse; so one can aim at making
them vanish or at minimizing them by shaping the pulses skillfully.
This is the route that we established previously 
\cite{pasin08a} analytically by the expansion in $\tau_p$. 
In the present work, we demonstrate numerically that the higher order 
corrections neglected  in the analytical calculations are indeed
negligible. Thereby, we have shown not only the validity of the
previous analytic calculation but we have also demonstrated that the real 
performance of the proposed pulses is advantageous.

We draw the readers' attention to the fact that shaped pulses have
 been introduced in NMR previously, see for instance
the Refs.\ 
\onlinecite{tycko83,geen91,cummi00,cummi03,chen01,garan02,motto06}
and Ref.\ \onlinecite{vande04} for an overview in the field
of quantum information. But the goals of these investigations
were different from ours even though it turned out that for
$\pi$ pulses certain shapes with $\tau_s=\tau_p/2$ happen to coincide 
\cite{pasin08a}.

The paper is organized as follows. In Sect.\ \ref{part1} we briefly
recall the analytical arguments for the expansion in powers 
of $\tau_p$; especially the expressions for the first and second order
corrections are given. Then we introduce 
a quantity to measure the deviation of the real pulse from
the ideal pulse and compute this deviation analytically.
Moreover, we relate the two parameters of the model to 
the experimental situation in various realizations of qubits.
In Sect. \ref{part2} the spin Hamiltonian is
introduced which serves as our system of a qubit coupled to a decoherence
bath.  For this model we compute the deviation between real and ideal
pulse analytically and numerically.  The experimentally relevant
ranges of parameters are estimated. The numerical results are discussed 
in Sect. \ref{part3} for $\pi$ and for $\pi/2$ pulses. Finally, in Sect.\ 
\ref{conclusions} we draw our conclusions.

\section{Theoretical predictions}
\label{part1}

\subsection{First and second order corrections}

In order to disentangle the actual pulse and the
free evolution of the system we proceed as follows.
The total unitary time evolution during the real pulse is
split into the time evolution of the system alone and of the
pulse alone which is taken
to occur at $\tau_s$ within the
interval $[0,\tau_p]$, see Ref.\ \onlinecite{pasin08a}. 
The time evolution of the system alone
is taken to occur before and after the evolution due to 
the pulse. The evolution due to the pulse is multiplied additionally
by corrections coming from the non-commutation of the 
Hamiltonians of the pulse and of the system. They can be expanded
in a series in $\tau_p$. 
It is important to stress that this technique does not aim at 
eliminating the coupling between the qubit and the bath completely,
but only at separating the effect of the pulse from that of bath.
The coupling between the qubit and the bath
remains active during the free evolution of the system.

To be explicit, we consider the following general Hamiltonian
\begin{equation}
\label{hamilt_gen} 
H_\text{tot}=H+H_0(t),
\end{equation} where the Hamiltonian $H$ of the qubit 
coupled to the bath is 
\begin{equation}
\label{hamilt_gen_bath} 
H=H_b + \lambda A \sigma_z,
\end{equation} 
where $H_b$ is a completely general bath and $A$ a completely
general  coupling operator acting on the bath. The Pauli matrices
represent operators acting on the qubit.
The internal energy scale of $H_b$ 
shall be denoted by $\omega_b$ while $\lambda$ is
the coupling constant between qubit and bath. 

Note that we assume only a coupling along the $z$ direction.
Hence the model contains only dephasing, i.e., a finite $T_2$.
No spin flips are possible so that $T_1=\infty$. Though
this represents a restriction it is well justified for
large magnetic fields along $z$ so that all other couplings
average out in the rotating-frame approximation.

The Hamiltonian of the pulse is denoted by $H_0$
\begin{equation}
\label{hamilt_pulse} 
H_0(t)= v(t) \sigma_y,
\end{equation} 
representing a rotation around the  $y$ axis. The pulse shape is given by 
the function $v(t)$. Note that $H_0$ and $H$  do not commute implying that
the unitary time evolution $U(\tau_p,0)$ during the application of a pulse
is a  non-trivial quantity.
 
Splitting the time evolution $U(\tau_p,0)$ into the time
evolutions during two intervals, $U(\tau_p,\tau_s)$ and 
$U(\tau_s,0)$, and  formally solving the Schr\"odinger equation
for each of them with a suitable ansatz we eventually obtain
(for details see Refs.\ \onlinecite{pasin08a})
\begin{eqnarray}
\nonumber
U_p(\tau_p,0) &=&  \mathrm{T}\left\{e^{-i\int_0^{\tau_p}H_{tot}(t)dt}
\right\}
\\ \nonumber
&=&e^{-i(\tau_p-\tau_s) H} e^{-i\sigma_y \int_{\tau_s}^{\tau_p}
 v(t) dt} U_F(\tau_p,0)
\\
&& e^{-i\sigma_y
\int_{0}^{\tau_s}v(t)dt} e^{-i \tau_s H}.
\label{U_p_alltogether}
\end{eqnarray}
where $U_F(\tau_p,0)$ represents the correction term. Without any
correction, i.e., for $U_F(\tau_p,0)=1$, the two exponentials
of the pulse can be combined in the middle of the right hand side
of Eq.\ \eqref{U_p_alltogether} so that the unitary operator
of the ideal pulse occurs
\begin{equation}
U_p(\tau_p,0) = e^{-i(\tau_p-\tau_s) H} e^{-i\sigma_y \int_{0}^{\tau_p}\ 
 v(t) dt} e^{-i \tau_s H}.
\end{equation}

\renewcommand{\arraystretch}{1.2}
\begin{table}[t]
\begin{center}
 \begin{tabular}{|c@{\extracolsep{0.1\columnwidth}}|c|c|c|}
\hline
$\tau_s$ & amplitude(s) & $\tau_i$ & $\eta^{(2)}$\\
\hline
\hline
\multicolumn{4}{|c|}{\textbf{SGLPi}}\\
\hline
$1/2$�����& $\pi/2$ �����& --��& --\\
\hline
\multicolumn{4}{|c|}{\textbf{UPi}}\\
\hline
$1/2$          & $\pm 7 \pi/6$      & $1/7$��      �& $0.04401$\\
�                  &��              & $6/7$���      & $0$\\
������������       &����������������&��             & $0.12295$\\

\hline
\multicolumn{4}{|c|}{\textbf{ASYPi}}\\
\hline
$0.34085$ ������      &  $\pm 13\pi/6$   & $3/4$�& $-0.00653$\\
�����                 &���              �&�������& $-0.14783$\\
���������������       &������������������&�������& $\phantom{-}0.18087$\\
\hline
\end{tabular}
\caption{Overview of the $\pi$ pulses implemented in the simulations. UPi 
and 
SGLPi are symmetric pulses ($\tau_s=\tau_p/2$). The switching 
instants $\tau_i$ and the amplitudes are given in units of $\tau_p$ and 
$1/\tau_p$, respectively. 
The column $\eta^{(2)}$ refers from top to bottom to the coefficients
$\eta_{21}$,  $\eta_{22}$, $\eta_{23}$ in units of
$\tau_p^2$, see Eq.\ \eqref{eq:eta-def} 
\label{tab_pulses_pi}}
\end{center}
\end{table}
\renewcommand{\arraystretch}{1}

The correction is expanded in a series in powers of $\tau_p$
\begin{equation}
U_F(\tau_p,0) = \exp(-i(\eta^{(1)}+\eta^{(2)}+\ldots))
\end{equation}
 where $\eta^{(j)}$ is the term of order $\tau_p^j$. We obtained
\cite{pasin08a}
\begin{subequations}
\label{eq:eta-def} 
\begin{eqnarray}
\label{eta1}
\eta^{(1)} &=&  (\eta_{11}\sigma_x+\eta_{12}\sigma_z)\lambda A\\
\label{eta2} 
\eta^{(2)} &=&
i\left(\eta_{21}\sigma_x + \eta_{22}\sigma_z\right)\lambda [H_b,A]
+ \eta_{23} \sigma_y \lambda^2 A^2.\qquad
\end{eqnarray} 
\end{subequations}
Note that $[H_b,A]$ is of the order of $\omega_b$ so that the 
corresponding term is indeed of order $\lambda\omega_b \tau_p^2$, thus
of second order in $\tau_p^2$. This becomes manifest in the
explicit integral equations for the coefficients $\eta_{ij}$
\begin{subequations}
\label{eta_ij}
\begin{eqnarray}
\label{eta11}
\eta_{11}&=& (\tau_p-\tau_s)\sin\psi_{\tau_p}+\tau_s\sin\psi_0
-\!\!\int_0^{\tau_p}\!\!\!\sin\psi_{t}\ dt\\
\label{eta12}
\eta_{12}&=& (\tau_p-\tau_s)\cos\psi_{\tau_p}+\tau_s\cos\psi_0
-\!\!\int_0^{\tau_p}\!\!\!\cos\psi_{t}\ dt\qquad\\
\eta_{21} &=& \frac{(\tau_p-\tau_s)^2}{2}\sin\psi_{\tau_p}
-\frac{\tau_s^2}{2}\sin\psi_0
\nonumber\\
&& -\int_0^{\tau_p}\Delta t\sin\psi_t\ dt
\label{eta21}
\\
\eta_{22} &=& -\frac{(\tau_p-\tau_s)^2}{2}\cos\psi_{\tau_p}
+\frac{\tau_s^2}{2}\cos\psi_0
\nonumber\\
&& +\int_0^{\tau_p}\Delta t\cos\psi_t\ dt
\label{eta22}
\\
\eta_{23} &=& (\tau_p-\tau_s)\tau_s\sin\theta
-\tau_s\int_0^{\tau_p}\sin(\psi_t-\psi_0)\ dt
\nonumber\\
&& -(\tau_p-\tau_s)\int_0^{\tau_p}\sin(\psi_{\tau_p}-\psi_t)\ dt
\nonumber\\
&&
+\frac{1}{2} \iint_0^{\tau_p}
\sin(\psi_{t_1}-\psi_{t_2})\text{sgn}(t_1-t_2)dt_1dt_2,\qquad
\label{eta23}
\end{eqnarray}
\end{subequations}
where $\psi_t=2 \int_{\tau_s}^t v(t') dt'$, $\Delta t=t-\tau_s$, and 
$\theta=\psi_{\tau_p}-\psi_0$ is the area under 
the amplitude of the pulse. The angle $\theta$ represents
the total angle of rotation of the qubit's spin under the action of the 
pulse. 

The function $v(t)$ and the instant $\tau_s$ are the free variables 
which can be fine-tuned to ideally make  the coefficients $\eta_{ij}$
vanish or at least to minimize their moduli.
In Fig.\ \ref{pulsesPi} examples of piecewise constant 
pulses for $\theta=\pi$ are reported. The pulse
SGLPi is the standard pulse of constant amplitude which
has finite first and second order corrections. The pulses  UPi and ASYPi
are chosen such that their first order correction $\eta^{(1)}$
vanishes. Their second order correction $\eta^{(2)}$ does not vanish.
We have proven previously that $\eta^{(2)}$ cannot be made vanish
for a $\pi$ pulse.
\cite{pasin08a}. For quantitative details, see Tab.\ \ref{tab_pulses_pi}.

\renewcommand{\arraystretch}{1.2}
\begin{table}[ht]
\begin{center}
\begin{tabular}{|c@{\extracolsep{0.1\columnwidth}}c|c|c|}
\hline
$\tau_s$ & amplitude(s) & $\tau_i$ & $\eta^{(2)}$\\
\hline
\hline
\multicolumn{4}{|c|}{\textbf{SGLPi2}}\\
\hline
$1/2$�����& $\pi/4$�����������������& --������������& --\\
\hline
\multicolumn{4}{|c|}{\textbf{UPi2}}\\
\hline
$1/2$�����  & $\pm 1.65765$�������& $0.13155$�����& $-0.01305$\\
������������&���������������������& $0.86845$�����& $0$\\
������������&���������������������&���������������& $\phantom{-}0.05151$\\

\hline
\multicolumn{4}{|c|}{\textbf{ASYPi2}}\\
\hline
$0.23128$�����& $\pm 1.39116$�����& $0.78220$�����& $-0.01279$\\
��������������&������������������ &���������������& $-0.05691$\\
��������������&�������������������&�������������� & $\phantom{-}0.88990$\\
\hline
\multicolumn{4}{|c|}{\textbf{S2NDPi2}}\\
\hline
$1/2$����      �& $\pm 2.31993$����������& $0.05848$�����& $0$\\
����������������&������������������������& $0.22384$�����& $0$\\
����������������&������������������������& $0.77616$�����& $\pm 0.01335$\\
����������������&������������������������& $0.94152$�����& \\
\hline
\multicolumn{4}{|c|}{\textbf{A2NDPi2}}\\
\hline
$0.61218$������     �& $\pm 2.09429$���& $0.08361$�����& $0$\\
���������������������&���������������� & $0.29828$�����& $0$\\
���������������������&�����������������& $0.90217$��� �& $\pm 0.01659$\\
\hline
\end{tabular}
\caption{Overview of the $\pi/2$ pulses implemented in the simulation. 
UPi2, SGLPi2, and S2NDPi2 are symmetric pulses ($\tau_s=\tau_p/2$). The 
switching instants $\tau_i$ and the amplitudes are given in units of 
$\tau_p$ and $1/\tau_p$, respectively. The column $\eta^{(2)}$ refers 
from top to bottom to the coefficients $\eta_{21}$,  $\eta_{22}$, 
$\eta_{23}$ in units of $\tau_p^2$, see Eq.\ 
\eqref{eq:eta-def}   
\label{tab_pulses_pi_halb}} 
\end{center}
\end{table}
\renewcommand{\arraystretch}{1}

In analogy, Fig.\ \ref{pulsesPi2} depicts examples of piecewise constant 
pulses for $\theta=\pi/2$. The pulse SGLPi2 is the standard pulse of 
constant amplitude which has finite first and second order corrections. 
The pulses UPi2 and ASYPi2 are chosen such that their first order 
correction $\eta^{(1)}$
vanishes. Their second order correction $\eta^{(2)}$ does not vanish.
For the quantitative details, we refer the reader
to Tab.\ \ref{tab_pulses_pi_halb}.

The pulses S2ND2 and A2NDPi2 are plotted in Fig.\ \ref{pulsesPi2nd}. 
They are chosen such that their first order correction 
$\eta^{(1)}$ \emph{and} the second order coefficients $\eta_{21}$
and $\eta_{22}$ vanish. We were not able to find a solution which
has additionally $\eta_{23}=0$. But we have not succeeded in proving
the impossibility of finding such a solution either.
For the quantitative details, we refer the reader
to Tab.\ \ref{tab_pulses_pi_halb}.

\begin{figure}
\begin{center}
     \includegraphics[width=0.9\columnwidth]{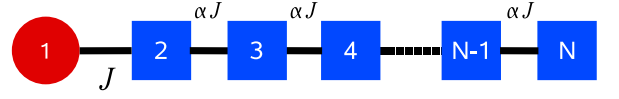}
\end{center}
\caption{(Color online)
Sketch of the spin chain representing qubit and spin bath. 
The qubit 1 is coupled to the spin 2 
of the chain. The coupling between qubit and the spin bath is 
given by $J$ while the internal exchange coupling within the chain
is $\alpha J$.}
\label{chain}
\end{figure}

\subsection{Measure of deviation}

The above results represent the analytical finding that we intend to check
numerically. In order to do so we need a measure of how well the real 
pulse approximates the ideal instantaneous one. We define the operator 
difference $\Delta:=U_p^\text{i}-U_p^\text{r}$ which quantifies the 
distance of the ideal time evolution ($U_p^\text{i}$) from the 
real one ($U_p^\text{r}$). To capture this distance by a single 
number we define the norm
\begin{eqnarray}
\label{d}
d &:=&\sqrt{\max\left\{\text {Eigenvalues}(\Delta^\dagger\Delta)\right\}}.
\end{eqnarray} 
For a pulse of angle $\theta$, the ideal pulse reads
\begin{equation}
 \label{U_ip}
U^\text{i}_p= 
e^{-i (\tau_p-\tau_s)H}e^{-i\frac{\theta}{2}\sigma_y}e^{-i\tau_s H}
\end{equation}
while the real pulse is given by
\begin{equation}
 \label{U_rp}
U^\text{r}_p= e^{-i (\tau_p-\tau_s)H}e^{-i\frac{\psi_{\tau_p}}{2}\sigma_y}
U_F(\tau_p,0)e^{i\frac{\psi_0}{2}\sigma_y}e^{-i\tau_s H},
\end{equation} 
such that
\begin{equation}
\label{delta_U_irp}
\Delta = e^{-i (\tau_p-\tau_s)H}
e^{-i\frac{\psi_{\tau_p}}{2}\sigma_y}(\mathbbm{1}-U_F(\tau_p,0))
e^{i\frac{\psi_0}{2}\sigma_y}e^{-i\tau_s H}.
\end{equation} 
This implies
\begin{equation}
 \label{delta_delta} \Delta^\dagger \Delta = \tilde{U}^\dagger 
(\mathbbm{1}-U_F^\dagger(\tau_p,0))(\mathbbm{1}-U_F(\tau_p,0))\tilde{U},
\end{equation}
where $\tilde{U}=e^{-i\frac{\psi_0}{2}\sigma_y}e^{-i\tau_s H}$ 
is a unitary operator which leaves the eigenvalues of the
product $(\mathbbm{1}-U_F^\dagger(\tau_p,0))(\mathbbm{1}-U_F(\tau_p,0))$ 
unaffected. 

Next, we  expand $U_F$ in $\tau_p$. If the leading order
is $\eta^{(1)}$ we have $U_F\approx \mathbbm{1}-i\eta^{(1)} +
{\cal O}(\tau_p^2)$ whence 
\begin{equation}
\label{d_eigenvalues1} 
d = 
\sqrt{\max \left \{ \text {Eigenvalues}( {\eta^{(1)}} \eta^{(1)}   ) 
\right \}+{\cal O}(\tau_p^3)}.
\end{equation}
If the leading order
is $\eta^{(2)}$ we have $U_F\approx \mathbbm{1}-i\eta^{(2)} +
{\cal O}(\tau_p^3)$ whence 
\begin{equation}
\label{d_eigenvalues2} 
d = \sqrt{\max \left \{ \text {Eigenvalues}( {\eta^{(2)}} \eta^{(2)}   ) 
\right \}+{\cal O}(\tau_p^5)}.
\end{equation}
We deduce that in the case of finite first order $\eta^{(1)}\neq 0$
one has $d= {\cal O}(\tau_p)$ while for vanishing first order,
but finite second order $\eta^{(2)}\neq 0$
one has $d= {\cal O}(\tau_p^2)$.

\section{The spin chain as decoherence bath}
\label{part2}

\subsection{The Model}

The formulae (\ref{d_eigenvalues1},\ref{d_eigenvalues2})  
for $d$  hold for any Hamiltonian that can 
be expressed in the form (\ref{hamilt_gen_bath}). 
Next, we specify the  model we investigate numerically. 
It is a spin chain of $N$  spins
where the first spin ($\widetilde{\sigma}$) represents the qubit,
see Fig.\ \ref{chain}. The  Hamiltonian considered is given by
\begin{equation}
\label{spin_Ham}
H_s=J \widetilde{\sigma}_z \sigma_z^{(2)}+\alpha J \sum_{i=2}^N 
\vec{\sigma}^{(i)}\cdot\vec{\sigma}^{(i+1)}.
\end{equation} 
Obviously, this Hamiltonian is an example for the most general
dephasing Hamiltonian \eqref{hamilt_gen_bath}.
In \eqref{spin_Ham} the bath is a bath of spins and the coupling between 
bath and qubit is quantified by $J$; hence we have $\lambda=J$.
The internal energy scale of the bath $\omega_b$ equals $\alpha J$
in \eqref{spin_Ham}.

In order to apply our general results 
(\ref{d_eigenvalues1},\ref{d_eigenvalues2})
we have to compute $\eta^{(2)}$ for the specific case of the spin 
Hamiltonian \eqref{spin_Ham}. 
The bath operator $A$ in (\ref{hamilt_gen_bath}) consists
only of the $z$ component of the second spin. Hence one has
$A^2=\mathbbm{1}$. 

The other term in \eqref{eta2} comprises $\left[H_b,A\right]$.
For \eqref{spin_Ham} this commutator contains only the second and the 
third spin. Hence we anticipate that the numerical results will not
show any significant size dependence in the regime where the
expansion in $\tau_p$ is valid, i.e., for low values of 
$\lambda$ and $\omega_b$ which translates to low values of
$J$. Explicitly we find for $\eta^{(2)}$
\begin{equation}
\label{eta2_spin}
\eta^{(2)}=-2J^2\alpha (\eta_{21}\widetilde{\sigma}_x+
\eta_{22}\widetilde{\sigma}_z)(\vec{\sigma}^{(2)}
\times\vec{\sigma}^{(3)})_z + \eta_{23}J^2\widetilde{\sigma}_y,
\end{equation} 
where $(\ )_z$ stands for the $z$ component. Because only three spins 
occur, it is a basic exercise to determine for $\eta^{(2)}$  given by 
\eqref{eta2_spin} the maximum eigenvalues of $(\eta^{(2)})^2$ yielding
\begin{equation}
\label{d_final}
d=J^2\sqrt{16\ \alpha^2(\eta_{21}^2+\eta_{22}^2)+\eta_{23}^2+ 
{\cal O}(\tau_p^5)}.
\end{equation}
In this formula, the quadratic dependence of $d$ as a function of 
$J$ has been put in evidence. The quadratic dependence on $\tau_p$
is less manifest, but it becomes obvious on inspecting
the integrals in \eqref{eta_ij} from which $\eta_{2j}={\cal O}(\tau_p^2)$
ensues.

Once $\tau_s$ and  $v(t)$ are known, the coefficients 
$\eta_{21}$, $\eta_{22}$, and $\eta_{23}$ 
can be easily computed according to (\ref{eta_ij}). 
Thereby, we have an analytical prediction for the leading order
of $d$ as function of $J$ including the prefactor.
 For fixed value of $J$, Eq.\
(\ref{d_final}) as a function of $\alpha$ is characterized by a 
constant behaviour dominated by $\eta_{23}$ for
$\alpha\ll 1$ and a linear behaviour in $\alpha$
for large values of the coupling constant.

\subsection{The Range of Parameters}

Although we are focusing here on the theoretical issues
it is helpful to have an idea about the experimental range of
parameters. In the sequel, we thus  try to assess the relevant
ranges. The numbers given represent only crude estimates
since the precise values depend strongly on the particular
experimental setup. Moreover, the relevant decoherence processes
are not yet always known.

First, we consider liquid NMR like crotonic acid or alanine 
\cite{fortu02}. The pulse lengths $\tau_p$ used are in the range of
$200\mu$s. The maximum pulse amplitude $B_m$ for a $\pi$ pulse
is thus in the range of 10kHz. The couplings between the nuclear
spins lie between 1 and about 70 Hz. A key ratio is $J/B_m$, i.e.,
the relative dimensionless strength of the pulse. Here it takes values 
in the range of $10^{-4}$ and $10^{-2}$. The other important 
parameter $\alpha$
is the dimensionless ratio $\omega_b/\lambda$ between the internal 
energy scale $\omega_b$ of the bath and the coupling between 
qubit and bath. Because the coupling between
the switched spin is typically of the same order  
as the coupling between the other spins $\alpha$ is roughly
of the order of 1.

Second, we consider a solid NMR system, namely KPF$_6$.
There, we found $B_m\approx 90$kHz and interspin couplings
ranging from $3.3$kHz to $11$kHz \cite{lovri07}
. This implies $J/B_m \approx 
0.04 - 0.12$ whereas $\alpha$ ranges between $0.3$ and $3$.
Another system is adamantane, for which we assume
$B_m\approx 150$kHz and $J\approx 15$kHz so that $J/B_m\approx 0.1$.
The ratio $\alpha$ is again taken to be of the order of 1
\cite{lovri08a}.

Third, we consider the electronic spin in a quantum dot as 
the qubit. The experimental investigation of temperature
dependent spin relaxation has just started \cite{herna08}.
The pulses are very short ($\tau_p\approx 1$ps) which
implies for a $\pi$ pulse according to 
$B_m \tau_p /\hbar = \Theta/2= \pi/2$ the amplitude
$B_m\approx 1$meV. But it is much less clear which $\lambda$ or
$\alpha$ one should consider. In Ref.\ \onlinecite{herna08}
a bosonic bath with spectral density $J_\text{eff}(\omega)$ is considered.
Taking the Debye frequency $\omega_\text{D}=27.5$meV as upper cutoff
and deducing $J$ from
\begin{equation}
\label{J-def}
J^2= \int_0^{\omega_\text{D}} J_\text{eff}(\omega)d\omega
\end{equation}
one obtains $J \approx 0.3 - 20$eV which implies enormous
values for $J/B_m$ but small values for $\alpha=\omega_\text{D}/J$.

But closer inspection of the estimates for $T_2$ \cite{herna08} 
reveals that the above estimate is not the relevant one.
Rather the internal energy scale appears to be set by the
energy splitting $\Delta\approx 70\mu$eV of the two qubit states.
The characteristic coupling is found by restricting the integral in
\eqref{J-def}  to the interval $[0,\Delta]$. Then $J\approx 1 - 6$neV ensues
which implies $J/B_m\approx 10^{-6} -10^{-5}$ and
$\alpha\approx 10^4$. Hernandez et al.\ \cite{herna08} doubt the relevance
of the spin relaxation via Rashba and Dresselhaus terms
advocating phonon-induced dephasing \cite{semen04,semen07}.
Then one should rather estimate $J^2\approx \Gamma\Delta$
with $\Gamma \approx 0.2\mu$eV implying $J\approx 4\mu$eV.
Then $J/B_m\approx0.004$ and $\alpha\approx 20$.
This example illustrates that the unambiguous identification
of the relevant processes of decoherence is still a challenging
task.

Fourth, we consider a qubit realized by
 charge states in a superconducting  device \cite{nakam02}.
The pulse length is $\tau_p\approx 80$ps implying 
$B_m\approx 15\mu$eV. The coupling $J$ is taken from
the free decay $J\approx \hbar/150\text{ps}\approx 5\mu$eV
while we deduce $\omega_b\approx 0.2\mu$eV from the decay of the signal
with an echo pulse. So $J/B_m \approx 0.3$ and $\alpha\approx 0.04$.

Fifth and last, we look at trapped ions \cite{leibf03} for
which we found pulse lengths in the range of microseconds implying
$B_m\approx 1$MHz. The coupling to optical modes
takes values $J\approx 20 - 200$kHz so that $J/B_m\approx 0.02 - 0.2$.
Less obvious is the relevant internal energy scale $\omega_b$.
The energies of the optical modes in the cavities are fairly
high between $1$ and $40$Ghz so that $\alpha$ would range in the
order of $10^{6}$. Thus the question arises whether this is
really the relevant scale or whether the very fast modes average out
so that a much lower effective scale comes into play.

The above numbers provide a rough guideline in which range today's
experiments are done. Surely, more elaborate investigations
of the relevant decoherence mechanisms are called for.

\begin{figure}
\begin{center}
     \includegraphics[scale=0.7,angle=-90]{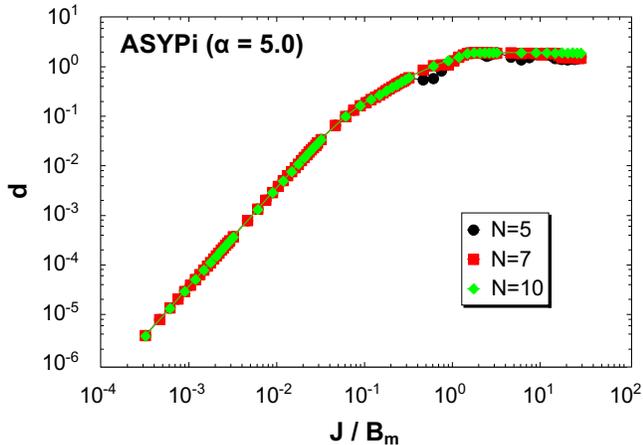}
\end{center}
\caption{(Color online)
Deviation $d$ as a function of $J/B_m$ ($B_m$ is the maximum
amplitude of the pulse) for various lengths $N$ of the spin chain at
$\alpha=5.0$. The data refers to the ASY1 pulse.}
\label{comp_spins}
\end{figure}
\begin{figure}
\begin{center}
     \includegraphics[width=\columnwidth]{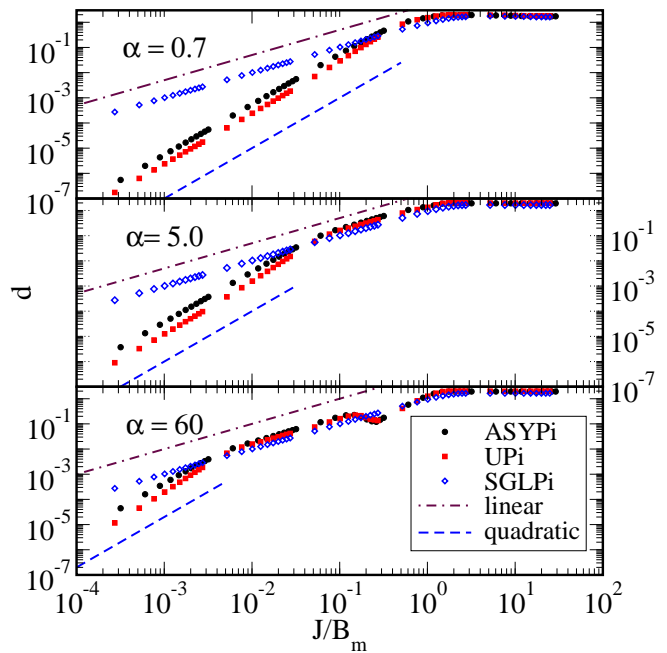}
\end{center}
\caption{(Color online) Case of $\pi$ pulses. The deviation $d$
 is plotted  as 
function of $J/B_m$ for $N=10$ and various values of $\alpha$.
For an unbiased comparison of the pulses, $J$ is normalized to
the maximum amplitude $B_m$ of the pulses. The notation for the pulses 
refers to Tab.\ \ref{tab_pulses_pi}. 
The dashed lines ease the comparison with pure power laws.}
\label{dvsJ_pi}
\end{figure}
\begin{figure}
\begin{center}
     \includegraphics[scale=0.65,angle=-90]{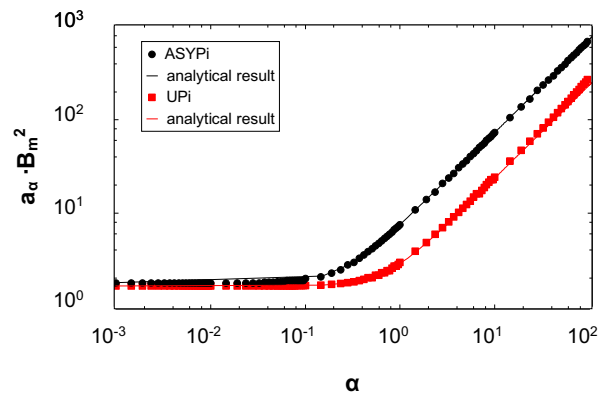}
\end{center}
\caption{(Color online)
Case of $\pi$ pulses. Plot of the prefactors $a_\alpha$ 
in $d=a_\alpha J^2+{\cal O}(J^3)$ for $N=7$. The solid lines represent the
analytical prediction in Eq.\ (\ref{d_final}).}
\label{slope_pi}
\end{figure}

\section{The numerical analysis}
\label{part3}

\paragraph*{Remarks on the program}
The numerical data was obtained using  C${++}$ routines.
Many of the matrix calculations were realized with the help of
the MATPACK-package \cite{gamme06}. The exponentials of the matrices were
calculated using routines adapted from  EXPOKIT \cite{sidje98} abbreviated
 \texttt{padm}. These are  techniques based on Pad\'{e} summation. 
Note that this approach is well-suited to deal with
piecewise constant pulses whereas continuously varying
pulses are not accessible.

As anticipated from the analytical calculation, only minor finite-size
effects occur. This is illustrated numerically in Fig.\ \ref{comp_spins}
for one particular pulse. But all other pulses show the same
behavior.
Indeed, the finite-size effects are completely negligible in the
region of small values of $J$. Hence we conclude that a moderate
number of bath spins is sufficient. In the data presented here we
routinely use $N=7$ and $N=10$. For these system sizes no particular
matrix algorithms are needed.

\paragraph*{$\pi$ pulses: vanishing linear  order}
We consider symmetric and asymmetric pulses with angle $\theta=\pi$
which satisfy $\eta_{11}=\eta_{12}=0$ as defined in (\ref{eta_ij}).
For comparison, also the standard pulse with constant amplitude
and finite $\eta^{(1)}$ is computed.

Fig.\ \ref{dvsJ_pi} shows the behavior of the deviations
$d$ as a function of $J/B_m$ for representative values of the parameter 
$\alpha$. Here $B_m$ is the maximum amplitude of the pulse.
At first thought, a plot as function of
$J\tau_p$ appears reasonable. But the comparison as function
of $J/B_m$ is fairer because the simple pulses, for instance
the standard one SGLPi, need only a smaller amplitude. Hence 
they can experimentally be realized with a shorter duration $\tau_p$
if the apparatus restricts the maximum applicable amplitude.
This advantage is accounted for by the plot versus $J/B_m$.

The quadratic behavior of ASYPi and UPi proves that the first order 
corrections are completely cancelled. This is not the case for SGLPi for 
which the numerical data display a linear  behavior for small $J$. 
For large values of $\alpha$, $d$ starts to deviate from 
the desired quadratic behavior even at relative small values of $J$. This
indicates that the internal energy scale $\omega_b=\alpha J$
becomes important.

The comparison between the standard pulse SGLPi and the optimized ones
ASYPi and UPi shows that a crossover takes place. For low values of
$J$ the pulses with vanishing first order outperform the standard pulse
due to their steeper decrease. At larger values of $J$ the more
complicated structure of the optimized pulses does not pay anymore
and SGLPi is slightly better.
Note that the value of $J$
where the crossover takes place depends on the value of $\alpha$.
For low values of $\alpha$ ASYPi and UPi pay up to much larger
values of $J$ than for large values of $\alpha$.

Data such as presented in Fig.\ \ref{dvsJ_pi} is used to determine
the prefactors $a_\alpha$ defined in
\begin{equation}
\label{eq:a-def}
d=a_\alpha J^2 + {\cal O}(J^3)
\end{equation}
by fits. The fits are made only within the range of 
validity of the quadratic behavior. 
The results are plotted in Fig.\ \ref{slope_pi}. They agree
perfectly with the analytical prediction from Eq.\ (\ref{d_final}).
For the quantitative comparison the coefficients $\eta_{21}$, 
$\eta_{22}$, and $\eta_{23}$ are explicitly computed for ASYPi and UPi2 
 by means of Eqs.\ (\ref{eta_ij}), see also Tab.\ \ref{tab_pulses_pi}.

\begin{figure}
\begin{center}
     \includegraphics[width=\columnwidth]{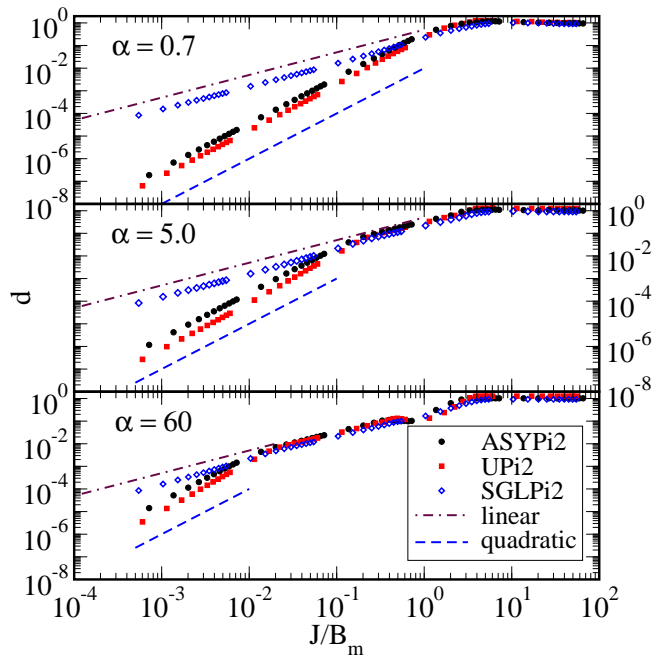}
\end{center}
\caption{(Color online)
Case of $\pi/2$ pulses. The deviation $d$
 is plotted  as function of $J/B_m$ for $N=10$ and various values of 
$\alpha$. For an unbiased comparison of the pulses, $J$ is normalized to
the maximum amplitude $B_m$ of the pulses. The notation for the pulses 
refers to Tab.\ \ref{tab_pulses_pi_halb}. 
The dashed lines ease the comparison with pure power laws.}
\label{dvsJ_pi_halb}
\end{figure}
\begin{figure}
\begin{center}
     \includegraphics[scale=0.70,angle=-90]{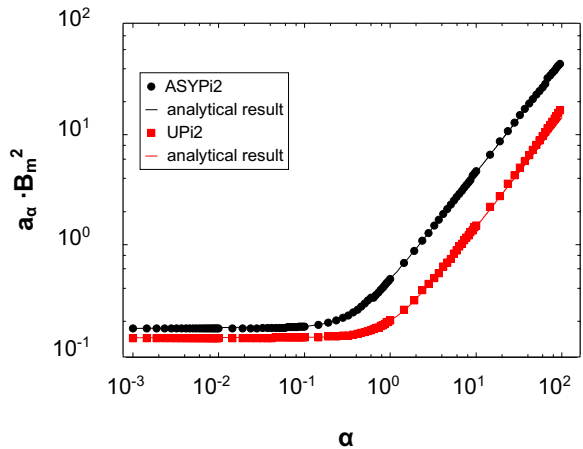}
\end{center}
\caption{(Color online)
Case of $\pi/2$ pulses. Plot of the prefactors $a_\alpha$ 
in $d=a_\alpha J^2+{\cal O}(J^3)$ for $N=7$. The solid lines represent the
analytical prediction in Eq.\ (\ref{d_final}).}
\label{slope_pi_halb}
\end{figure}

\paragraph*{$\pi/2$ pulses: vanishing linear order}
We consider symmetric and asymmetric pulses with angle $\theta=\pi/2$ 
which satisfy $\eta_{11}=\eta_{12}=0$ as defined in (\ref{eta_ij}).
For comparison, also the standard pulse with constant amplitude
and finite $\eta^{(1)}$ is computed.

Fig.\ \ref{dvsJ_pi_halb} shows the behavior of the deviations
$d$ as a function of $J/B_m$ for representative values of the parameter 
$\alpha$. Again, the comparison as function
of $J/B_m$ is fairer for the above mentioned reasons.

The quadratic behavior of ASYPi2 and UPi2 proves that the first order 
corrections are completely cancelled. This is not the case for SGLPi2 for 
which the numerical data displays a linear behavior for small $J$. 
For large values of $\alpha$, $d$ starts to deviate from 
the desired quadratic behavior even at relative small values of $J$. This
indicates that the internal energy scale $\omega_b=\alpha J$
becomes important.

The comparison between the standard pulse SGLPi2 and the optimized ones
ASYPi2 and UPi2 shows that a crossover takes place. For low values of
$J$ the pulses with vanishing first order outperform the standard pulse
due to their steeper decrease. At larger values of $J$ the more
complicated structure of the optimized pulses does not pay anymore
and SGLPi2 is slightly better. Note that the value of $J$
where the crossover takes place depends on the value of $\alpha$.
For low values of $\alpha$ ASYP2i and UPi2 pay up to much larger
values of $J$ than for large values of $\alpha$.

Note that the gain of the optimized pulses over the standard pulse
is most significant for low values of $\alpha$, i.e., for
a slow internal bath dynamics. It is less significant for 
fast internal bath dynamics corresponding to large values of $\alpha$.

Data such as presented in Fig.\ \ref{dvsJ_pi_halb} is used to determine
the prefactors $a_\alpha$ defined in \eqref{eq:a-def}
by fits. The fits are made only within the range of 
validity of the quadratic behavior. 
The results are plotted in Fig.\ \ref{slope_pi_halb}. They agree
perfectly with the analytical prediction in Eq.\ (\ref{d_final}).
For the quantitative comparison the coefficients $\eta_{21}$, 
$\eta_{22}$, and $\eta_{23}$ are explicitly computed for ASYPi2 and UPi2 
 by means of Eqs.\ (\ref{eta_ij}), see also Tab.\ 
\ref{tab_pulses_pi_halb}.

The errors in the fits of the prefactors for the $\pi/2$ pulses were
determined  by hand for some randomly chosen data points using
various fitting ranges and fitting functions such as $b J +
a_\alpha J^2$ or $a_\alpha J^2 + b J^3$. This analysis provides the error
estimates of 12\% at $\alpha=0.03$ to 32\% at $\alpha=28$ for pulse 
UPi2 and about 10\% for all $\alpha$ looking at pulse ASYPi2.

\begin{figure}
\begin{center}
     \includegraphics[width=\columnwidth]{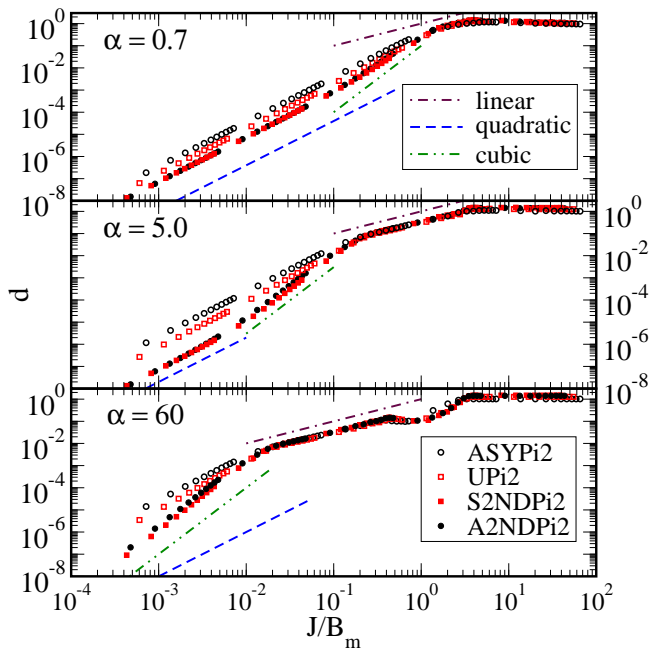}
\end{center}
\caption{(Color online)
Case of $\pi/2$ pulses with partly vanishing
quadratic order. The deviation $d$
 is plotted  as function of $J/B_m$ for $N=10$ and various values of 
$\alpha$. For an unbiased comparison of the pulses, $J$ is normalized to
the maximum amplitude $B_m$ of the pulses. The notation for the pulses 
refers to Tab.\ \ref{tab_pulses_pi_halb}. 
The dashed lines ease the comparison with pure power laws.}
\label{dvsJ_pi_halb_2ord}
\end{figure}

\paragraph*{$\pi/2$ pulses: partly vanishing quadratic order}
In the previous work in Refs.\ \cite{pasin08a} 
we have proven rigorously that no $\pi$ pulse can satisfy the 
second order Eqs.\ (\ref{eta21},\ref{eta22},\ref{eta23}). 
For $\pi/2$ pulses no such proof is known to us. But
we were not able to find a solution to all five equations
Eqs.\ (\ref{eta_ij}) either.

We managed, however, to find solutions which make the first
four equations (\ref{eta11},\ref{eta12},\ref{eta21},\ref{eta22})
vanish. The advantage is that the first order vanishes completely
and that in second order all the terms of order $\lambda\omega_b \tau_p^2$
vanish also. Only the term of order $\lambda^2 \tau_p^2$ persists.
We expect such pulses, see Fig.\ \ref{pulsesPi2nd} and 
Tab.\ \ref{tab_pulses_pi_halb}, to be advantageous for
systems where the coupling $\lambda$ between qubit and bath
is very small, but the internal bath dynamics $\omega_b$ is not.

Here we propose  two possible examples of $\pi /2$ pulses, 
symmetric and asymmetric, for which $\eta_{11}=\eta_{12}=0$,
$\eta_{21}=\eta_{22}=0$, but $\eta_{23}\neq 0$. From the 
above arguments, we expect that for large values of $\alpha$,
i.e., fairly fast baths, the deviation $d(J)$ displays cubic
behavior at least in some intermediate range.
Fig.\ \ref{dvsJ_pi_halb_2ord} provides the corresponding data.
Indeed, one clearly identifies an intermediate range where
cubic behavior is seen. This range is fairly small for small
values of $\alpha$ (upper panel in Fig.\ \ref{dvsJ_pi_halb_2ord})
but grows upon increasing $\alpha$
(middle panel in Fig.\ \ref{dvsJ_pi_halb_2ord}).
For the large values of $\alpha$ analyzed in the lower panel in 
Fig.\ \ref{dvsJ_pi_halb_2ord} the quadratic behavior below
the cubic range is not even discernible. But we know
from Eq.\ \eqref{d_final} that is exists.

We conclude that even a partial vanishing of the second
order can be very helpful. This conclusion is supported
by the comparison to data for ASYPi2 and UPi2 which
have a vanishing first order, but no vanishing second order
terms. As to be expected, we find that for low values of $J$
the pulses S2NDPi2 and A2NDPi2 outperform ASYPi2 and UPi2.
For larger values of $J$ a crossover takes place and 
there is no need to resort to the more complicated
pulses S2NDPi2 and A2NDPi2. There, all pulses behave
very much alike.

Note that the crossover takes place for lower values
of $J$ if $\alpha$ is large and viceversa for larger values of $J$
if $\alpha$ is small. This is related to the fact that
the range of cubic behavior occurs at larger values of $J$ 
for small $\alpha$. For large $\alpha$ the range is
larger, but shifted to smaller values of $J$.

\section{Conclusions}
\label{conclusions}

We  numerically simulated the effect of designed short control pulses 
on a qubit coupled to a bath of spins. 
The effect of the short pulse can be approximated in leading order of the 
pulse duration $\tau_p$ as a $\delta$ peak.
For finite  $\tau_p$, however, corrections occur which we
know from previous analytical calculations. The aim of the present work
was two-fold. First, we wanted to confirm the analytical results by
numerical calculations. Second, we intended to analyze to which extent
the analytically neglected higher orders matter. Put differently,
we wanted to see whether pulses, which are fine-tuned to make
the leading corrections vanish, outperform the standard pulses.

The numerical results confirm the analytical results in all
points. The fine-tuned pulses display qualitatively different
power laws in the deviation $d$ as function of $J\tau_p$.
This deviation measures the difference between the ideal pulse,
multiplied with the evolution due to 
free decoherence, and the realistic pulse.
For standard pulses, one has $d\propto J$. For the fine-tuned
pulses we achieve $d\propto J^2$.

In restricted parameter ranges, we obtained even $d\propto J^3$
for pulses which make certain parts of the second order corrections
vanish. Such pulses were not yet discussed before. They are only
possible for $\theta\neq\pi$.

The second goal has also been achieved by the analysis of the real 
performance in case of the coupling to a spin bath. 
We could show that the fine-tuned pulses outperform the more standard 
ones in a large range of parameters. Furthermore, we estimated
the relevant parameters for a number of generic experiments. 
These estimates show that many experimental setups are such
that the fine-tuned pulses should improve on the standard pulses.
But more investigations, both theoretical and experimental, are
needed to obtain a complete understanding of the important decoherence
mechanisms.

For the above reasons
we suggest that the choice of the optimized pulses with respect to
the standard ones is in many cases preferable. Our findings here
will provide guidelines under which experimental circumstances one should
use the optimized pulses.

\begin{acknowledgments}
We would like to thank  M.~Bayer, T.~Fischer, A.~Greilich, M.~Lovri\'c,
 and  J.~Stolze for helpful discussions. The financial support in GK 726 
by the DFG is gratefully acknowledged.
\end{acknowledgments}


\end{document}